\documentclass[12pt]{revtex4}

\usepackage{graphicx}
\usepackage{amsmath}
\usepackage{amssymb}
\usepackage{color}

\begin{document} 

\title
{Evidence from stable isotopes and $^{10}$Be for solar system formation triggered
by a low-mass supernova} 

\author{Projjwal Banerjee}
\affiliation{School of Physics and Astronomy, University of Minnesota, Minneapolis, 
Minnesota 55455, USA}

\author{Yong-Zhong Qian\footnote{Corresponding author.}\footnote{
Visiting Professor, Center for Nuclear Astrophysics, Department of Physics and 
Astronomy, Shanghai Jiao Tong University, Shanghai 200240, China.}}
\affiliation{School of Physics and Astronomy, University of Minnesota, Minneapolis, 
Minnesota 55455, USA; qian@physics.umn.edu}

\author{Alexander Heger\footnotemark[2]}
\affiliation{Monash Centre for Astrophysics, School of Physics and Astronomy, 
Monash University, Victoria 3800, Australia}

\author{W. C. Haxton}
\affiliation{Department of Physics, University of California, and Lawrence Berkeley 
National Laboratory, Berkeley, California 94720, USA}

\begin{abstract}
About 4.6 billion years ago, some event disturbed a cloud of gas and dust,
triggering the gravitational collapse that led to the formation of the solar system. 
A core-collapse supernova, whose shock wave is capable of 
compressing such a cloud, is an obvious candidate for the initiating event. 
This hypothesis can be tested because supernovae also produce telltale
patterns of short-lived radionuclides, which would be preserved today as isotopic 
anomalies. Previous studies of the forensic evidence have been inconclusive, 
finding a pattern of isotopes differing from that produced in conventional supernova 
models. Here we argue that these difficulties either do not arise or are mitigated 
if the initiating supernova was a special type, low in mass and explosion energy.  
Key to our conclusion is the demonstration that short-lived $^{10}$Be 
can be readily synthesized in such supernovae by neutrino interactions, while 
anomalies in stable isotopes are suppressed.
\end{abstract}

\maketitle 

Nearly four decades ago Cameron and Truran \cite{ct} suggested 
that the formation of our solar system (SS) might have been due 
to a single core-collapse supernova (CCSN) whose shock wave
triggered the collapse of a nearby interstellar cloud. They 
recognized that forensic evidence of such an event would be 
found in CCSN-associated short-lived ($\lesssim10$~Myr)
radionuclides (SLRs) that would decay, but leave a record of 
their existence in isotopic anomalies. Their suggestion was in fact 
stimulated by observed meteoritic excesses in $^{26}$Mg \cite{lee},
the daughter of the extinct SLR $^{26}$Al with a
lifetime of $\tau\sim 1$~Myr. The inferred value of 
$^{26}$Al/$^{27}$Al in the early SS, orders of 
magnitude higher than the Galactic background, 
requires a special source \cite{wasserburg}.

While simulations support the thesis that a CCSN shock wave can
trigger SS formation and inject SLRs into the early SS
\cite{boss2010,boss2014,boss2015},
detailed modeling of CCSN nucleosynthesis and an accumulation of 
data on extinct radionuclides have led to a confusing and conflicting
picture \cite{wasserburg,meyer}.
CCSNe of $\gtrsim 15$ solar masses ($M_\odot$) are a major source
of stable isotopes such as $^{24}$Mg, $^{28}$Si, and $^{40}$Ca.
The contributions from a single CCSN in this mass range combined
with the dilution factor indicated by simulations 
\cite{boss2010,boss2014,boss2015}
would have caused large shifts in ratios of stable isotopes 
that are not observed \cite{wasserburg}.
A second problem concerns the relative production of key SLRs:
such a CCSN source grossly overproduces $^{53}$Mn and $^{60}$Fe
\cite{wasserburg}, while producing (relatively) far too
little of $^{10}$Be. Although the overproduction of 
$^{53}$Mn and $^{60}$Fe can 
plausibly be mitigated by the fallback of inner CCSN material, 
preventing the ejection of these two SLRs \cite{meyer,takigawa},
the required fallback must be extremely efficient in high-mass 
CCSNe.

Here we show that the above difficulties with the CCSN trigger 
hypothesis can be removed or mitigated, if the CCSN mass
was $\lesssim 12\,M_\odot$. The structure of a low-mass CCSN
progenitor differs drastically from that of higher-mass counterparts,
being compact with much thinner processed shells.  Given the CCSN
trigger hypothesis, we argue that
the stable isotopes alone demand such a progenitor.
But in addition, this assumption addresses several other problems 
noted above. First, we show the yields of $^{53}$Mn and $^{60}$Fe
are reduced by an order of magnitude or more
in low-mass CCSNe, making the fallback required to bring 
the yields into agreement with the data
much more plausible. Second, we show that the 
mechanism by which CCSNe produce $^{10}$Be,
the neutrino spallation process 
$^{12}{\rm C}(\nu,\nu'pp)^{10}{\rm Be}$, 
differs from other SLR production mechanisms in that
the yield of $^{10}$Be remains high as the progenitor 
mass is decreased.  Consequently we
find that an $11.8\,M_\odot$ model can produce the bulk of the 
$^{10}$Be inventory in the early SS without 
overproducing other SLRs.  We conclude that among possible CCSN triggers, 
a low-mass one is demanded by the data on both 
stable isotopes and SLRs.

It has been commonly thought that $^{10}$Be is not associated with 
stellar sources, originating instead only from spallation of carbon and 
oxygen in the interstellar medium (ISM) by cosmic rays 
(CRs; e.g., \cite{desch}) or irradiation of the early SS material by
solar energetic particles (SEPs; e.g., \cite{gou1,gou2}) 
associated with activities of the proto-Sun. 
It was noted in Ref.~\cite{yoshida} that $^{10}$Be can be produced by 
neutrino interactions in CCSNe, but the result was presented for a single 
model and no connection to meteoritic data was made. Further, that work 
adopted an old rate for the destruction reaction 
$^{10}$Be($\alpha,n$)$^{13}$C that is orders of magnitude larger than 
currently recommended \cite{cyburt}, and therefore, greatly 
underestimated the $^{10}$Be yield. 

$^{10}$Be has been observed in the form of a $^{10}$B excess 
in a range of meteoritic samples. Significant variations 
across the samples suggest that multiple sources might have 
contributed to its inventory in the early SS
\cite{mckeegan,marhas,macpherson,liu,wielandt,srin}.
Calcium-aluminum-rich inclusions (CAIs) with $^{26}$Al/$^{27}$Al
close to the canonical value were found to have significantly higher
$^{10}$Be/$^9$Be than CAIs with Fractionation and Unidentified
Nuclear isotope effects (FUN-CAIs), which also have 
$^{26}$Al/$^{27}$Al much less than the canonical value \cite{wielandt}. 
As FUN-CAIs are thought to have formed earlier than canonical CAIs, 
it has been suggested \cite{wielandt} that the protosolar cloud was seeded 
with $^{10}$Be/$^9$Be~$\sim 3\times 10^{-4}$, the level observed in
FUN-CAIs, by e.g., trapping Galactic CRs \cite{desch}, and 
that the significantly higher $^{10}$Be/$^9$Be values in canonical CAIs 
were produced later by SEPs \cite{gou1,gou2}.
 
A recent study \cite{epi} showed that trapping Galactic CRs 
led to little $^{10}$Be enrichment of the protosolar cloud and long-term
production by Galactic CRs could only provide
$^{10}$Be/$^9$Be~$\lesssim 1.3\times 10^{-4}$. Instead, CRs from either
a large number of CCSNe or a single special CCSN were proposed to 
account for $^{10}$Be/$^9$Be~$\sim 3\times 10^{-4}$. 
While this pre-enrichment scenario is plausible, it depends on many 
details of CCSN remnant evolution and CR production and
interaction. Similarly, further production of $^{10}$Be by SEPs must have 
occurred at some level, but the actual contributions are sensitive to
the composition, spectra, and irradiation history of SEPs
as well as the composition of the irradiated gas and solids
\cite{gou1,gou2,duprat}, all of which are rather uncertain.
In view of both the data and uncertainties in CR and SEP models, 
we consider it reasonable that a low-mass CCSN provided 
the bulk of the $^{10}$Be inventory in the early SS while still allowing
significant contributions from CRs and SEPs. Specifically, we find
that such a CCSN can account for
$^{10}$Be/$^9$Be~$=(7.5\pm2.5)\times 10^{-4}$ typical of the 
canonical CAIs \cite{dauphas}. Following the presentation of our detailed 
results, we will discuss an overall scenario to account for $^{10}$Be and
other SLRs based on our proposed low-mass CCSN trigger and 
other sources.

\section*{Results}
\noindent
\underline{Explosion Modeling} \\
We have calculated CCSN nucleosynthesis for solar-composition progenitors
in the mass range of 11.8 to 30$\,M_\odot$.  Each star was evolved to core 
collapse, using the most recent version of the 1D hydrodynamic code KEPLER 
\cite{kepler1,kepler2}. The subsequent explosion was simulated by driving 
a piston from the base of the oxygen shell into the collapsing progenitor. 
Piston velocities were selected to produce
explosion energies of 0.1, 0.3, 0.6, and 1.2~B (1~B~$=10^{51}$~ergs)
for the 11.8--12, 14, 16, and 18--$30\,M_\odot$ models, respectively, 
to match results from recent CCSN simulations \cite{bruenn,melson}.
The material inside the initial radius of the piston was allowed to fall immediately 
onto the protoneutron star forming at the core. In our initial calculations, 
shown in Fig.~1 and labeled
Case 1 in Table~1, we assume all material outside the piston is ejected.  
Neutrino emission was modeled by assuming Fermi-Dirac spectra with 
chemical potentials $\mu=0$, fixed temperatures $T_{\nu_e} \sim 3$~MeV and
$T_{\bar\nu_e} \sim T_{\nu_\mu} \sim T_{\nu_\tau} \sim T_{\bar\nu_\mu} \sim 
T_{\bar\nu_\tau} \sim 5$~MeV, and luminosities decreasing exponentially from 
an initial value of 16.7~B~s$^{-1}$ per species, governed 
by a time constant of $\sim 3$~s. This treatment is consistent with detailed 
neutrino transport calculations \cite{nu_spectra3} as well as
supernova 1987A observations \cite{beacom2007}. 
A full reaction network was used to track changes in composition during the 
evolution and explosion of each star, including neutrino rates 
taken from Ref.~\cite{heger}.\\

\noindent
\underline{Nucleosynthesis Yields}\\
Figure~1 shows the yields normalized to the $11.8\,M_\odot$ model 
as functions of the progenitor mass for stable isotopes $^{12}$C, 
$^{16}$O, $^{24}$Mg, $^{28}$Si, $^{40}$Ca, and $^{56}$Fe as well as 
SLRs $^{10}$Be, $^{41}$Ca, $^{53}$Mn, $^{60}$Fe, and $^{107}$Pd.
It can be seen that except for $^{10}$Be, the yields of all other isotopes
increase sharply for CCSNe of 14--$30\,M_\odot$. Therefore, a high-mass
CCSN trigger is problematic, generating unacceptably large shifts in ratios of 
stable isotopes and overproducing SLRs such as 
$^{53}$Mn and $^{60}$Fe \cite{wasserburg}. 
Fallback of $\gtrsim 1\,M_\odot$ of inner material in such CCSNe was invoked 
in Ref.~\cite{takigawa} to account for the data on the SLRs $^{26}$Al, 
$^{41}$Ca, $^{53}$Mn, and $^{60}$Fe. Using our models (see Supplementary
Table~1), we find that similar fallback scenarios and dilution factors are 
required but the problem with stable isotopes 
persists (see Supplementary Discussion). In contrast, even for Case 1 
without fallback, the yields of the $11.8\,M_\odot$ model (see Supplementary
Tables~2 and 3) are consistent 
with meteoritic constraints for all major stable isotopes 
(see Supplementary Discussion). We focus on the production of SLRs 
by this model below.

Figure~1 shows that in contrast to other isotopes, the $^{10}$Be yield from 
$^{12}$C via $^{12}{\rm C}(\nu,\nu'pp)^{10}{\rm Be}$ is relatively insensitive to 
progenitor mass. This reflects the compensating effects of higher C-zone 
masses but lower neutrino fluxes (larger C-zone radii) in more massive stars
(see Supplementary Discussion for more on SLR production). 
Our demonstration here that $^{10}$Be is a ubiquitous CCSN product of
neutrino-induced nucleosynthesis consequently allows us to attribute
this SLR to a low-mass CCSN, explaining its abundance level
in canonical CAIs, while achieving overall consistency with the data on 
other SLRs coproduced by other mechanisms in the CCSN.
More quantitatively, let $R$ denote a given SLR, $I$ its stable 
reference isotope, $Y_\mathrm{R}$ the total mass yield of $R$ from the CCSN, 
and $f$ the fraction of the yield that was incorporated into each $M_\odot$ 
of the protosolar cloud (i.e., the dilution factor).  
The number ratio of $R$ to $I$ in the early SS due to this CCSN is
\begin{equation}
\left(\frac{N_\mathrm{R}}{N_\mathrm{I}}\right)_{\rm ESS}\sim
\frac{f Y_\mathrm{R}/A_\mathrm{R}}{X_\mathrm{I}^\odot M_\odot/A_\mathrm{I}} \exp\left(-\frac{\Delta}{\tau_\mathrm{R}}\right),
\end{equation}
where $A_\mathrm{R}$ and $A_\mathrm{I}$ are the mass numbers of $R$ and $I$,
$X_\mathrm{I}^\odot$ is the solar mass fraction of $I$ \cite{asplund}, 
$\Delta$ is the time between the CCSN explosion and incorporation of $R$ 
into early SS solids, and $\tau_\mathrm{R}$ is the lifetime of $R$.

Table~1 gives the mass yields of $^{10}$Be, $^{26}$Al, $^{36}$Cl, $^{41}$Ca,
$^{53}$Mn, $^{60}$Fe, $^{107}$Pd, $^{135}$Cs, $^{182}$Hf, and 
$^{205}$Pb for the $11.8\,M_\odot$ model.
A comparison of equation~(1) to the observed value, including uncertainties
\cite{dauphas,davis,al26,lin,hsu,jacobsen,ca1,ca2,mn53,tang,mishra,pd107,cs135,hf182,pb1,pb2}, 
yields a  band of allowed $f$ and $\Delta$ for each SLR. 
Simultaneous explanation of SLRs then requires the corresponding bands to 
overlap. Figure~2 shows a region of concordance for $^{10}$Be, $^{41}$Ca, 
and $^{107}$Pd. This fixes $f$ and $\Delta$, allowing us to estimate the 
contributions from the $11.8\,M_\odot$ CCSN to other SLRs. The Case 1 
contributions to $^{26}$Al, $^{36}$Cl, $^{53}$Mn, $^{60}$Fe, $^{135}$Cs, 
$^{182}$Hf, and $^{205}$Pb in Table~1 correspond to $f \sim 5\times 10^{-4}$ 
and $\Delta \sim 1$~Myr, the approximate best-fit point indicated by the filled 
circle in Fig.~2.

The slow-neutron-capture ($s$) process product $^{182}$Hf is of special interest, 
as the yield of this SLR is sensitive to the $\beta$-decay rate of $^{181}$Hf, which 
may be affected by thermally-populated low-lying excited states under stellar 
conditions. We treat the excited-state contribution as an uncertainty \cite{lugaro2014}, 
allowing the rate to vary between the laboratory value and the theoretical estimate of 
Ref.~\cite{hf181} with excited states. (The latter is numerically close to updated
estimates with uncertainties \cite{lugaro2014}.) 
The yield obtained with the laboratory rate 
accounts for almost all of the $^{182}$Hf in the early SS. This removes a conflict 
with data on the SLR $^{129}$I that arises when $^{182}$Hf is attributed to the 
rapid neutron-capture ($r$) process \cite{lugaro2014,wasserburg2}. \\

\noindent
\underline{Role of Fallback}\\
The Case~1 results of Table~1 are consistent
with the meteoritic data on $^{26}$Al, $^{36}$Cl, $^{135}$Cs, 
$^{182}$Hf, and $^{205}$Pb, as the contributions do not exceed the measured
values.  In contrast, although the production of $^{53}$Mn and $^{60}$Fe is 
greatly reduced in low-mass CCSNe, the $^{53}$Mn contribution remains 
a factor of 60 too large while $^{60}$Fe is compatible only with the
larger of the two observed values (see Table~1).  Both of 
these SLRs originate from zones deep within the $11.8\,M_\odot$ star:
$^{53}$Mn is produced in the innermost $10^{-2}\,M_\odot$ of the shocked
material, while $\sim 90$\% of the $^{60}$Fe is associated with the 
innermost $0.12\,M_\odot$. Because of the low explosion energy 
used here based on simulations \cite{melson},
the expected fallback of the innermost shocked zones onto the 
protoneutron star \cite{zhang} provides a natural explanation for the 
discrepancies: most of the produced $^{53}$Mn and, possibly, $^{60}$Fe 
is not ejected. In Case~2 of Table~1, where only 1.5\% of the innermost 
$1.02\times 10^{-2}\,M_\odot$ is ejected, $^{53}{\rm Mn}/^{55}{\rm Mn}$ is 
reduced to its measured value $(6.28\pm0.66)\times 10^{-6}$ \cite{mn53},
while other SLR contributions are largely unaffected. In Case 3, where only 1.5\%
of the innermost $0.116\,M_\odot$ is ejected, additional large reductions 
(a factor of $\sim 10$) are found for $^{60}$Fe and $^{182}$Hf, 
accompanied by smaller decreases (a factor of $\sim 2$) in $^{26}$Al, $^{36}$Cl, 
$^{135}$Cs, and $^{205}$Pb. 

Case 3 represents the limit of reducing $^{53}$Mn and $^{60}$Fe without 
affecting the concordance among $^{10}$Be, $^{41}$Ca, and $^{107}$Pd 
(see Supplementary Fig.~1, Supplementary Discussion).
Were the lower observed value for $^{60}$Fe \cite{tang} proven correct, 
we would have to either reduce its yield by examining the
significant nuclear and stellar physics uncertainties \cite{limongi,woosley} or
use even more substantial fallback and reconsider the low-mass CCSN 
contributions to SLRs. Because of the correlated effects of fallback on $^{60}$Fe 
and $^{182}$Hf, more fallback would also rule out an attractive explanation for 
the latter, as described above.  
Note that the fallback assumed for Cases 2 and 3 is far below that invoked
for high-mass CCSNe in Ref.~\cite{takigawa} to account for $^{26}$Al, 
$^{41}$Ca, $^{53}$Mn, and the higher observed value of $^{60}$Fe.

If, however, the higher $^{60}$Fe value \cite{mishra} is correct, then a plausible 
scenario like Case 2, where SS formation was triggered by a low-mass CCSN 
with modest fallback, would be in reasonable agreement with the data on
$^{10}$Be, $^{41}$Ca,  $^{53}$Mn, $^{60}$Fe, and $^{107}$Pd. The nuclear
forensics, notably the rapidly decaying $^{41}$Ca, determines the delay between 
the CCSN explosion and incorporation of SLRs into early SS solids, 
$\Delta \sim 1$~Myr. 
The deduced fraction of CCSN material injected into the protosolar cloud, 
$f \sim 5\times 10^{-4}$, is consistent with estimates based on simulations 
of ejecta interacting with dense gas clouds \cite{boss2010,boss2014,boss2015}
(see Supplementary Discussion). There is also an 
implicit connection to the CCSN explosion energy, which influences fallback in 
hydrodynamic models.

\section*{Discussion}
In addition to neutrino-induced production, a low-mass CCSN can make $^{10}$Be
through CRs associated with its remnant evolution \cite{epi}. However, the yield of 
this second source is modest (see Supplementary Discussion). The net yield in the 
ISM trapped within the remnant is limited by the amount of this ISM.
Production within the general protosolar cloud during its initial contact with the 
remnant (i.e., prior to thorough mixing of the injected material) would also be 
expected, and the yield could possibly account for 
$^{10}{\rm Be}/^9{\rm Be}\sim 3 \times 10^{-4}$ in FUN-CAIs \cite{epi}.
However, FUN-CAIs are rare, and their $^{10}$Be inventory may be more 
consistent with local production by the CCSN CRs. Taking the net CR 
contribution averaged over the protosolar cloud to be 
$^{10}{\rm Be}/^{9}{\rm Be}\sim 10^{-4}$, a value that we argue is more 
consistent with long-term production by Galactic CRs \cite{epi}, we add
the neutrino-produced $^{10}{\rm Be}/^{9}{\rm Be}\sim(5.2$--$6.4)\times 10^{-4}$ 
(see Table~1) from the CCSN to obtain
$^{10}{\rm Be}/^{9}{\rm Be}\sim(6.2$--$7.4)\times 10^{-4}$, which is in
accord with $^{10}{\rm Be}/^{9}{\rm Be}=(7.5\pm2.5)\times 10^{-4}$ observed
in canonical CAIs. In general, we consider that neutrino-induced production
provided the baseline $^{10}$Be inventory in these samples and the observed
variations \cite{mckeegan,macpherson,wielandt,srin} can be largely attributed 
to local production by SEPs.

Our proposal that a low-mass CCSN trigger provided the bulk of the $^{10}$Be 
inventory in the early SS has several important features: (1) the relevant 
neutrino and CCSN physics is known reasonably well, and the uncertainty 
in the $^{10}$Be yield is estimated here to be within a factor of $\sim 2$;
(2) the production of both $^{10}$Be and $^{41}$Ca is in agreement 
with observations \cite{ca1,ca2}, a result difficult to achieve by SEPs 
\cite{srin}; and (3) the yield pattern of Li, Be, and B isotopes
(see Supplementary Table~4) is distinctive,
with predominant production of $^7$Li and $^{11}$B and 
differing greatly from patterns of production by CRs and SEPs, so that 
precise meteoritic data might provide distinguishing tests 
(see Supplementary Discussion).

We emphasize that while $^{53}$Mn and $^{60}$Fe production is greatly
reduced in a low-mass CCSN, some fallback is still required to explain the
meteoritic data. The fallback solution works well for $^{53}$Mn (see Table~1). 
When somewhat different meteoritic values of $^{53}$Mn/$^{55}$Mn 
\cite{mn53a,mn53b} are used, only the ejected fractions of the innermost 
shocked material need to be adjusted accordingly. 
The case of $^{60}$Fe is more complicated.
The meteoritic measurements are difficult, especially in view of a recent 
study showing the mobility of Fe and Ni in the relevant samples \cite{telus1}. 
Another recent study gave
$5\times 10^{-8}\lesssim{^{60}{\rm Fe}}/^{56}{\rm Fe}\lesssim 2.6\times 10^{-7}$
\cite{telus2}, which may be accounted for by Case 3 of our model (see Table~1).
However, were ${^{60}{\rm Fe}}/^{56}{\rm Fe}\sim 10^{-8}$ \cite{tang}, currently 
preferred by many workers, to be confirmed, we would have to conclude that 
either the present $^{60}$Fe yield of the low-mass CCSN is wrong or its 
contributions to SLRs must be reconsidered. 

Several other issues with our proposed low-mass CCSN trigger merit discussion.
Table~1 shows that such a CCSN underproduces $^{26}$Al, $^{36}$Cl, and 
$^{135}$Cs to varying degrees. We consider that the ISM swept up by the 
CCSN shock wave prior to triggering the collapse of the protosolar cloud might 
have been enriched with $^{26}$Al by nearby massive stars. To avoid
complications with $^{53}$Mn and $^{60}$Fe, we propose that these
stars might have exploded only weakly or not at all \cite{zhang}, but contributed
$^{26}$Al through their winds. The total amount of swept-up $^{26}$Al needed 
to be $\sim 10^{-5}\,M_\odot$ (see Table 2), which could have been provided by 
winds from stars of $\gtrsim 35\,M_\odot$ \cite{limongi}, possibly in connection 
with an evolving giant molecular cloud \cite{vasil}. Winds from massive stars
may also have contributed to $^{41}$Ca and $^{135}$Cs \cite{wind}. 
However, the wind contribution to $^{41}$Ca might be neglected given the rapid 
decay of this SLR over the interval of $\sim 1$~Myr between the onset of collapse 
of the protosolar cloud and incorporation of SLRs into early SS solids 
(see Supplementary Discussion).
We agree with previous studies that $^{36}$Cl was probably produced by SEPs 
after most of the initial $^{26}$Al had decayed \cite{hsu,jacobsen}. 
The corresponding late irradiation would not have caused problematic 
coproduction of other SLRs, especially $^{10}$Be, $^{26}$Al, and $^{53}$Mn, 
if it occurred in a reservoir enriched with volatile elements such as chlorine, 
a major target for producing $^{36}$Cl \cite{jacobsen}.

Our calculations do not include
nucleosynthesis in the neutrino-heated ejecta from the 
protoneutron star, where some form of the $r$ process may take place
\cite{woosley2,wanajo}. This is a potential source of the SLR $^{129}$I.
As emphasized above, a low-mass CCSN would alter the SS ratios of 
stable isotopes of e.g., Mg, Si, Ca, and Fe only at levels of 
$\lesssim 1\%$ (see Supplementary Discussion), 
consistent with meteoritic constraints \cite{wasserburg}. Nonetheless, 
Cases 2 and 3 with fallback would produce anomalies in $^{54}$Cr, 
$^{58}$Fe, and $^{64}$Ni at levels of $\sim 10^{-3}$ as observed in 
meteorites (see Supplementary Discussion). As there are few satisfactory 
explanations of these anomalies \cite{wass}, this provides circumstantial 
support for the fallback scenario required by the $^{53}$Mn and $^{60}$Fe data.

We conclude that a low-mass CCSN is a promising 
trigger for SS formation. Such a trigger is plausible because
the lifetime of $\sim 20$~Myr for the CCSN progenitor is compatible with 
the duration of star formation in giant molecular clouds \cite{gmc}.
Further progress depends on resolving
discrepancies in $^{60}$Fe abundance determinations, clarifying 
the nuclear physics of $^{181}$Hf decay, and studying the evolution 
of additional low-mass CCSN progenitors and their explosion,
especially quantifying fallback through multi-dimensional models.
In addition, the overall scenario proposed here to explain the SLRs 
in the early SS requires comprehensive modeling of $^{26}$Al
enrichment by winds from massive 
stars in an evolving giant molecular cloud, evolution of a low-mass CCSN
remnant and the resulting CR production and interaction, and irradiation
by SEPs associated with activities of the proto-Sun.
Finally, tests of the low-mass
CCSN trigger by precise measurements of Li, Be, and B isotopes in 
meteorites are highly desirable (see Supplementary Discussion).

\section*{Data Availability}
The data that support the findings of this study are available from the 
corresponding author upon reasonable request.

\section*{Acknowledgements}
We acknowledge helpful discussions with Bernhard M\"uller and 
the late Jerry Wasserburg. We thank Takashi Yoshida for 
communications regarding Ref.~\cite{yoshida}.
This work was supported in part by 
the US DOE [DE-FG02-87ER40328 (UM), 
DE-SC00046548 (Berkeley), and DE-AC02-98CH10886 (LBL)],
the US NSF [PHY-1430152 (JINA-CEE)],
and ARC Future Fellowship FT120100363 (AH).

\section*{Author contributions}
P.B. and Y.-Z.Q. designed the work. P.B. ran the models with help from A.H.
All the authors discussed the results and contributed to the writing of the 
manuscript.

\section*{Competing financial interests}
The authors declare no competing financial interests.

\clearpage
\begin{figure}
\centerline{\includegraphics[width=5in]{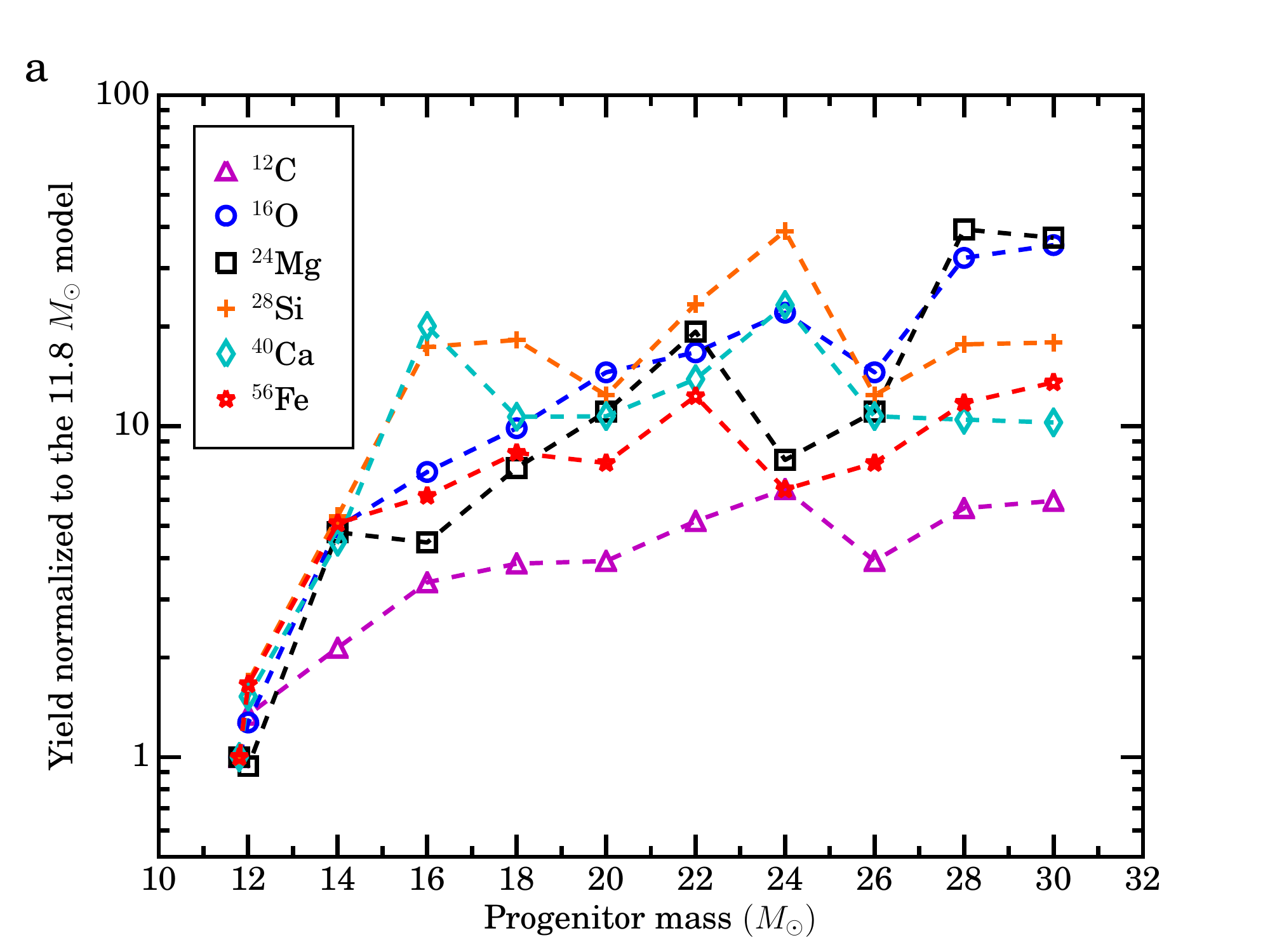}}
\centerline{\includegraphics[width=5in]{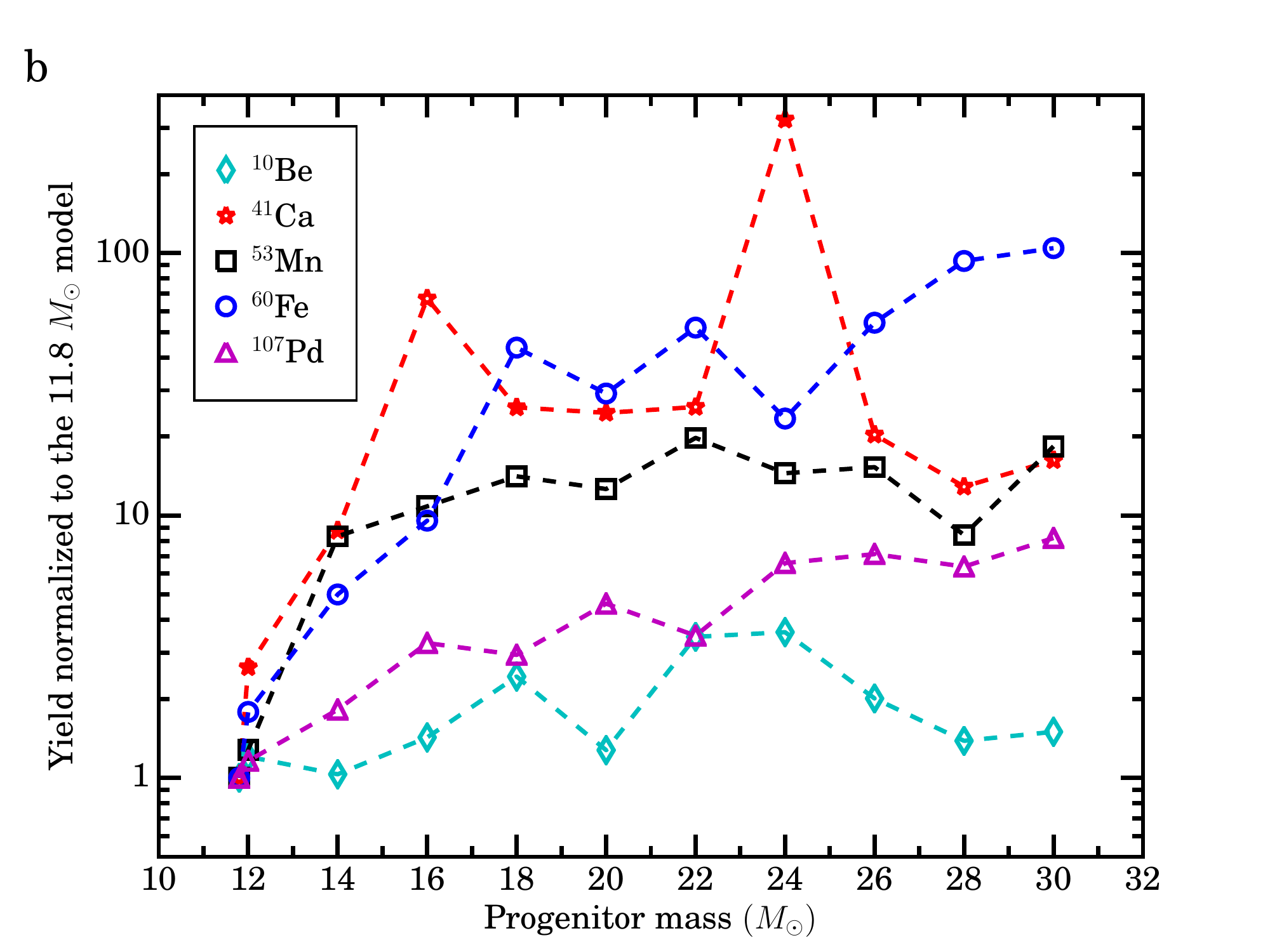}}
\caption{Nucleosynthetic yields as functions of the supernova progenitor's mass.  
Selected yields of (a) stable isotopes and (b) short-lived radionuclides are shown, 
normalized to the 11.8-solar-mass model, for Case 1 with no fallback.}
\end{figure}

\begin{figure}
\centerline{\includegraphics[width=6in]{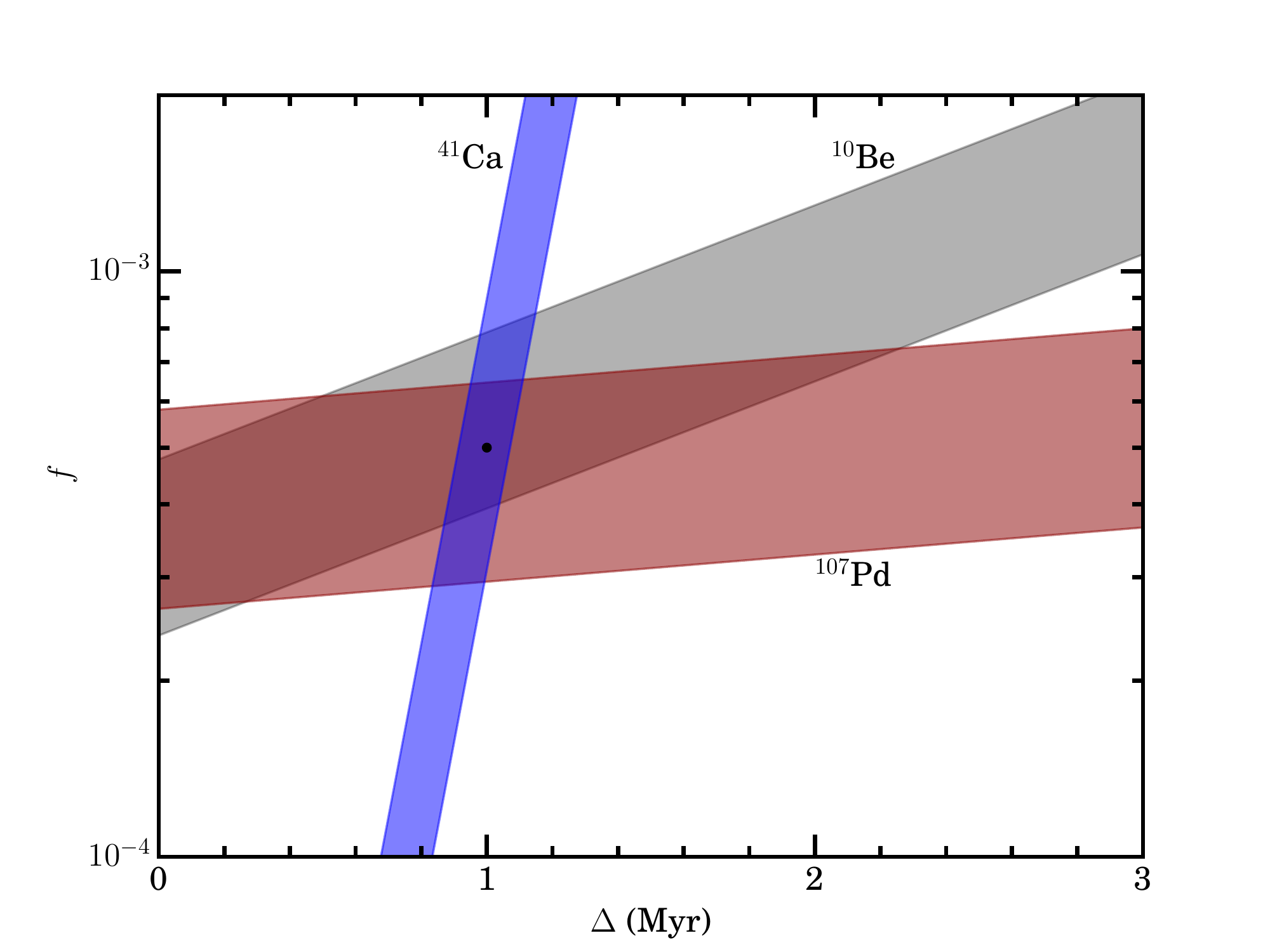}}
\caption{Relations between parameters characterizing the core-collapse supernova trigger.
The parameter $f$ denotes the fraction of the yields of short-lived nuclides incorporated 
into the proto-solar cloud, per solar mass.  The parameter $\Delta$ denotes
the time between the supernova explosion and incorporation of short-lived nuclides into 
early solar system solids. Results are calculated from equation~(1) using yields for the
11.8-solar-mass model with no fallback (Case 1) and
meteoritic data for $^{10}$Be, $^{41}$Ca, and $^{107}$Pd with
$2\sigma$ uncertainties (see Table~1). The filled circle at 
$f \sim 5\times 10^{-4}$ and $\Delta \sim 1$~Myr is the approximate best-fit 
point within the overlap region.}
\end{figure}

\clearpage
\begin{table}
Table 1: Yields of short-lived radionuclides from an 11.8-solar-mass core-collapse supernova

~\\
\centerline{\begin{tabular}{|c|c|c|c|c|c|c|c|}
\hline
$R/I$ &$\tau_\mathrm{R}$  &$Y_\mathrm{R}$& $X_\mathrm{I}^\odot$&\multicolumn{4}{|c|} {$~~(N_\mathrm{R}/N_\mathrm{I})_{\rm ESS}$} \\
\cline{5-8}
&(Myr)&$(M_\odot)$&&Data&Case 1&Case 2&Case 3\\
\hline
\rule{0pt}{2ex}${\rm ^{10}{Be}/^{9}{Be}}$&2.00&$3.26(-10)$&$1.40(-10)$&{$(7.5\pm 2.5)(-4)$}&$6.35(-4)$&$6.35(-4)$ &$5.20(-4)$\\
\rule{0pt}{2ex}${\rm ^{26}Al/^{27}Al}$&1.03&$2.91(-6)$&$5.65(-5)$&$(5.23\pm 0.13)(-5)$&$1.02(-5)$&$9.90(-6)$&$5.77(-6)$\\
\rule{0pt}{2ex}${\rm ^{36}Cl/^{35}Cl}$&0.434&$1.44(-7)$&$3.50(-6)$&$\sim(3$--$20)(-6)$&$2.00(-6)$&$1.45(-6)$&$6.15(-7)$\\
\rule{0pt}{2ex}${\rm ^{41}{Ca}/^{40}{Ca}}$&0.147&$3.66(-7)$&$5.88(-5)$&{$(4.1\pm 2.0)(-9)$}&$3.40(-9)$&$2.74(-9)$&$2.26(-9)$\\
\rule{0pt}{2ex}${\rm ^{53}Mn/^{55}Mn}$&5.40&$1.22(-5)$&$1.29(-5)$&$(6.28\pm0.66)(-6)$ &$4.04(-4)$&$6.39(-6)$&$6.16(-6)$\\
\rule{0pt}{2ex}${\rm ^{60}Fe/^{56}Fe}$&3.78&$3.08(-6)$&$1.12(-3)$&$\sim 1(-8);(5$--$10)(-7)$&$9.80(-7)$&$9.80(-7)$&$1.10(-7)$\\
\rule{0pt}{2ex}${\rm ^{107}{Pd}/^{108}{Pd}}$&9.38&$1.37(-10)$&$9.92(-10)$&{$(5.9\pm 2.2)(-5)$}&$6.27(-5)$&$6.27(-5)$&$5.72(-5)$\\
\rule{0pt}{2ex}${\rm ^{135}Cs/^{133}Cs}$&3.32&$2.56(-10)$&$1.24(-9)$&$\sim 5(-4)$&$7.51(-5)$ &$7.51(-5)$&$3.18(-5)$\\
\rule{0pt}{2ex}${\rm ^{182}Hf/^{180}Hf}$&12.84&$4.04(-11)$&$2.52(-10)$&$(9.72\pm 0.44)(-5)$&$7.36(-5)$&$7.36(-5)$&$6.34(-6)$\\
&&$8.84(-12)$&&&$1.60(-5)$&$1.60(-5)$&$2.37(-6)$\\
\rule{0pt}{2ex}${\rm ^{205}Pb/^{204}Pb}$&24.96&$9.20(-11)$&$3.47(-10)$&$\sim 1(-4);1(-3)$&$1.27(-4)$&$1.27(-4)$&$7.78(-5)$\\
\hline
\end{tabular}}
\end{table}

\noindent Comparisons are made to the corresponding isotopic ratios 
deduced from meteoritic data. 
Case 1 estimates are calculated from equation~(1) 
using the approximate best-fit $f$ and 
$\Delta$ of Fig. 2, assuming no fallback.
The higher and lower yields for $^{182}$Hf are obtained from the laboratory 
and estimated stellar decay rates \cite{hf181} of $^{181}$Hf, respectively.
Case 2 (3) is a fallback scenario in which only 1.5\% of the innermost 
$1.02\times10^{-2}$ solar mass (0.116 solar mass) of shocked 
material is ejected. With guidance from Refs.~\cite{dauphas,davis}, 
well-determined data are quoted with $2\sigma$ errors,
while data with large uncertainties are preceded by ``$\sim$''.
Note that $x(-y)$ denotes $x\times 10^{-y}$. Data references are:
$^{10}$Be \cite{mckeegan,macpherson,wielandt,srin},
$^{26}$Al \cite{lee,al26},
$^{36}$Cl \cite{lin,hsu,jacobsen},
$^{41}$Ca \cite{ca1,ca2},
$^{53}$Mn \cite{mn53},
$^{60}$Fe \cite{tang,mishra},
$^{107}$Pd \cite{pd107},
$^{135}$Cs \cite{cs135},
$^{182}$Hf \cite{hf182},
$^{205}$Pb \cite{pb1,pb2}.

\clearpage
\setcounter{equation}{0}
\setcounter{figure}{0}
\setcounter{table}{0}
\makeatletter
\renewcommand{\theequation}{S\arabic{equation}}
\renewcommand{\thefigure}{S\arabic{figure}}
\renewcommand{\bibnumfmt}[1]{[S#1]}
\renewcommand{\citenumfont}[1]{S#1}

\begin{figure}
\centerline{\includegraphics[width=5in]{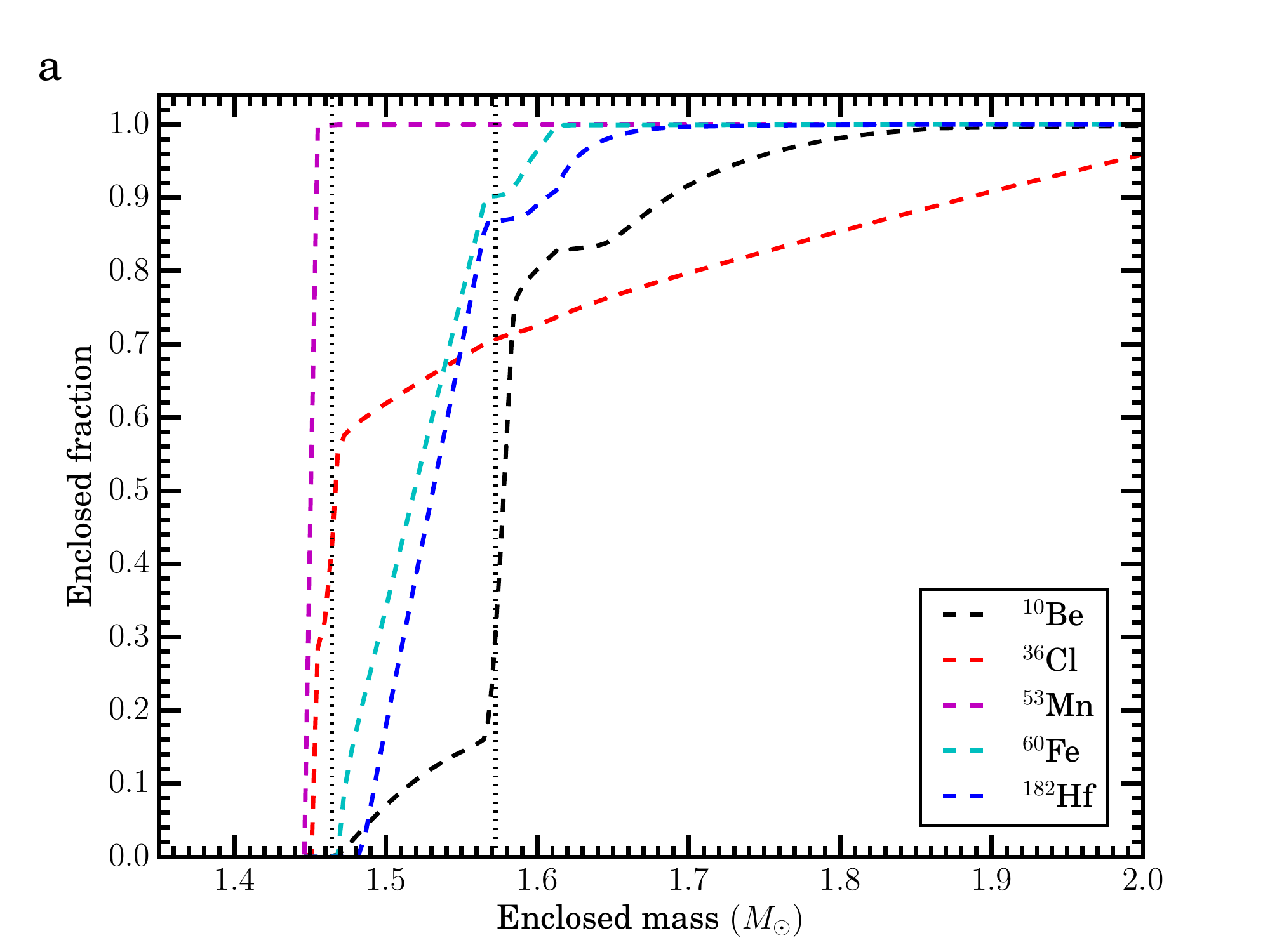}}
\centerline{\includegraphics[width=5in]{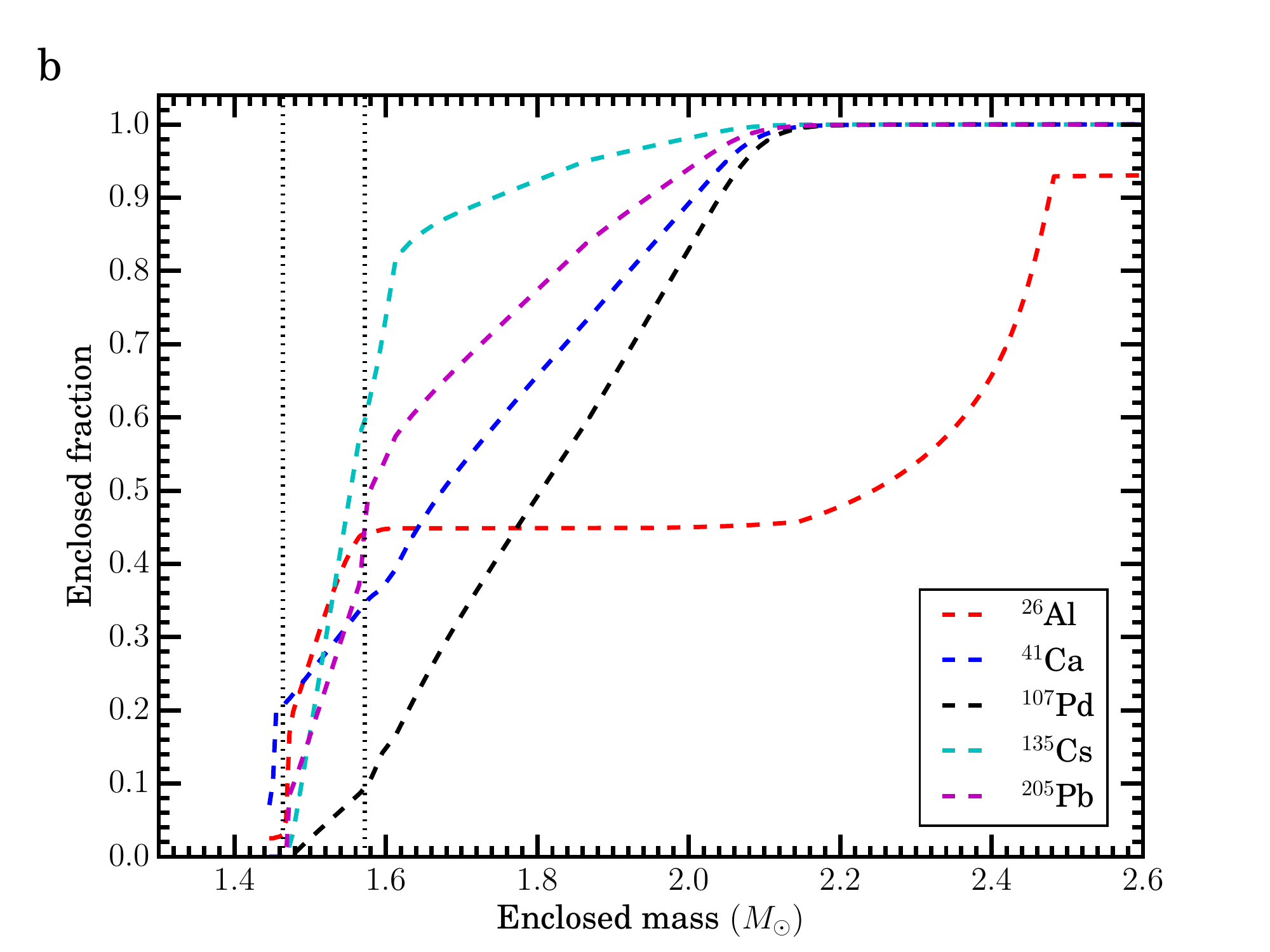}}
\end{figure}
\vskip 0.5cm
\noindent Supplementary Figure~1: Enclosed fraction of the total yield of 
each short-lived radionuclide
as a function of the enclosed stellar mass for the 11.8-solar-mass model 
with no fallback (Case 1). The left and right vertical lines indicate the 
boundary between fallback and ejected material for Cases 2 and 3, 
respectively.

\clearpage
\begin{table}
Supplementary Table~1: Supernova contributions to short-lived radionuclides 
in the early solar system from the 20- and 30-solar-mass models
\vskip 0.5cm
\centerline{\begin{tabular}{|c|c|c|}
\hline
$R/I$&$20\,M_\odot$&$30\,M_\odot$\\
\hline 
\rule{0pt}{2ex}${\rm ^{10}{Be}/^{9}{Be}}$&$7.39(-5)$&$3.70(-5)$\\  
\rule{0pt}{2ex}${\rm ^{26}Al/^{27}Al}$&$5.64(-5)$&$5.17(-5)$\\  
\rule{0pt}{2ex}${\rm ^{36}Cl/^{35}Cl}$&$2.89(-6)$&$1.37(-6)$\\  
\rule{0pt}{2ex}${\rm ^{41}{Ca}/^{40}{Ca}}$&$4.95(-9)$&$4.21(-9)$\\
\rule{0pt}{2ex}${\rm ^{53}Mn/^{55}Mn}$&$6.24(-6)$&$6.40(-6)$\\
\rule{0pt}{2ex}${\rm ^{60}Fe/^{56}Fe}$&$1.04(-6)$&$9.48(-7)$\\   
\rule{0pt}{2ex}${\rm ^{107}{Pd}/^{108}{Pd}}$&$3.02(-4)$&$1.47(-4)$\\    
\rule{0pt}{2ex}${\rm ^{135}Cs/^{133}Cs}$&$3.01(-4)$&$5.85(-5)$\\
\rule{0pt}{2ex}${\rm ^{182}Hf/^{180}Hf}$&$4.54(-5)$&$7.52(-6)$\\
\rule{0pt}{2ex}${\rm ^{205}Pb/^{204}Pb}$&$4.76(-04)$&$1.35(-4)$\\
\hline
\end{tabular}}
\end{table}

\noindent Fallback scenarios similar to those of Ref.~\cite{stakigawa}
are assumed.
Values for $^{182}$Hf are obtained from the estimated stellar
decay rate of $^{181}$Hf \cite{shf181}. Note that $^{107}$Pd 
is overproduced and $x(-y)$ denotes $x\times 10^{-y}$.

\clearpage
\begin{table}
Supplementary Table~2: Yields of major stable isotopes for the 11.8-solar-mass model
with no fallback (Case 1)
\vskip 0.5cm
\centerline{\begin{tabular}{|l|c|l|c|l|c|}
\hline
 Isotope&Yield $(M_\odot)$&Isotope&Yield $(M_\odot)$&Isotope&Yield $(M_\odot)$\\
\hline 
\rule{0pt}{2ex}$^{12}$C   &$5.75(-2)$&$^{33}$S&$3.39(-5)$&$^{54}$Cr&$5.08(-6)$\\
\rule{0pt}{2ex}$^{13}$C   &$6.98(-4)$&$^{34}$S&$1.91(-4)$&$^{54}$Fe&$8.58(-4)$\\
\rule{0pt}{2ex}$^{14}$N   &$2.68(-2)$&$^{36}$S&$1.18(-6)$&$^{56}$Fe&$1.77(-2)$\\
\rule{0pt}{2ex}$^{15}$N   &$1.96(-5)$&$^{40}$Ca&$7.28(-4)$&$^{57}$Fe&$5.51(-4)$\\
\rule{0pt}{2ex}$^{16}$O   &$1.43(-1)$&$^{42}$Ca&$4.00(-6)$&$^{58}$Fe&$9.23(-5)$\\
\rule{0pt}{2ex}$^{17}$O   &$5.41(-5)$&$^{43}$Ca&$1.05(-6)$&$^{58}$Ni&$6.95(-4)$\\
\rule{0pt}{2ex}$^{18}$O   &$7.46(-4)$&$^{44}$Ca&$1.64(-5)$&$^{60}$Ni&$3.53(-4)$\\
\rule{0pt}{2ex}$^{20}$Ne &$3.55(-2)$&$^{46}$Ca&$5.36(-8)$&$^{61}$Ni&$2.76(-5)$\\
\rule{0pt}{2ex}$^{21}$Ne&$1.17(-4)$&$^{48}$Ca&$1.21(-6)$&$^{62}$Ni&$8.73(-5)$\\
\rule{0pt}{2ex}$^{22}$Ne&$3.16(-3)$&$^{46}$Ti &$2.22(-6)$&$^{64}$Ni&$2.00(-5)$\\
\rule{0pt}{2ex}$^{24}$Mg&$8.60(-3)$&$^{47}$Ti &$2.29(-6)$&$^{64}$Zn &$1.03(-5)$\\
\rule{0pt}{2ex}$^{25}$Mg&$1.50(-3)$&$^{48}$Ti &$3.03(-5)$&$^{66}$Zn &$8.75(-6)$\\
\rule{0pt}{2ex}$^{26}$Mg&$1.61(-3)$&$^{49}$Ti &$2.19(-6)$&$^{67}$Zn &$1.40(-6)$\\
\rule{0pt}{2ex}$^{28}$Si&$9.11(-3)$ &$^{50}$Ti  &$1.80(-6)$&$^{68}$Zn &$5.64(-6)$\\
\rule{0pt}{2ex}$^{29}$Si&$4.38(-4)$ &$^{50}$Cr&$9.32(-6)$ &$^{70}$Zn &$1.63(-7)$\\
\rule{0pt}{2ex}$^{30}$Si&$4.29(-4)$ &$^{52}$Cr&$2.15(-4)$ &&\\
\rule{0pt}{2ex}$^{32}$S&$4.71(-3)$ &$^{53}$Cr&$2.74(-5)$ &&\\
\hline
\end{tabular}}
\end{table}

\noindent Note that $x(-y)$ denotes $x\times 10^{-y}$.

\clearpage
\begin{table}
Supplementary Table~3: Percentage of the solar system inventory of major stable 
isotopes contributed by the 11.8-solar-mass model
\vskip 0.5cm
\centerline{\begin{tabular}{|l|c|c|c|l|c|c|c|}
\hline
Isotope&Case 1&Case 2&Case 3&Isotope&Case 1&Case 2& Case 3\\
\hline 
\rule{0pt}{2ex}$^{12}$C   &1.16   &1.16    &1.10&$^{48}$Ca&0.45  &0.45    &0.44\\
\rule{0pt}{2ex}$^{13}$C   &1.16   &1.16    &1.16&$^{46}$Ti&0.46  &0.46    &0.45\\
\rule{0pt}{2ex}$^{14}$N   &1.82   &1.82    &1.82&$^{47}$Ti&0.52  &0.45    &0.44\\
\rule{0pt}{2ex}$^{15}$N   &0.33   &0.33    &0.22&$^{48}$Ti&0.68  &0.44    &0.44\\
\rule{0pt}{2ex}$^{16}$O   &1.18  &1.18    &0.56&$^{49}$Ti&0.65  &0.51    &0.48\\
\rule{0pt}{2ex}$^{17}$O   &1.10  &1.10    &1.06&$^{50}$Ti&0.55  &0.55    &0.48\\
\rule{0pt}{2ex}$^{18}$O   &2.72  &2.72    &2.72&$^{50}$Cr&0.63  &0.44    &0.44\\
\rule{0pt}{2ex}$^{20}$Ne &1.44  &1.44    &0.48&$^{52}$Cr&0.73  &0.45     &0.44\\
\rule{0pt}{2ex}$^{21}$Ne&1.89   &1.89    &0.98&$^{53}$Cr&0.80  &0.45    &0.45\\
\rule{0pt}{2ex}$^{22}$Ne &1.59  &1.59    &1.56&$^{54}$Cr&0.59  &0.59    &0.50\\
\rule{0pt}{2ex}$^{24}$Mg &0.86  &0.86   &0.45&$^{54}$Fe&0.62  &0.44    &0.44\\
\rule{0pt}{2ex}$^{25}$Mg&1.14  &1.14    &0.51&$^{56}$Fe&0.79  &0.45    &0.44\\
\rule{0pt}{2ex}$^{26}$Mg&1.07  &1.07    &0.66&$^{57}$Fe&1.04  &0.50    &0.48\\
\rule{0pt}{2ex}$^{28}$Si  &0.70  &0.57    &0.44&$^{58}$Fe&1.29  &1.29    &0.73\\
\rule{0pt}{2ex}$^{29}$Si  &0.64  &0.64    &0.46&$^{58}$Ni&0.71  &0.44    &0.44\\
\rule{0pt}{2ex}$^{30}$Si  &0.92  &0.92    &0.47&$^{60}$Ni&0.91  &0.50    &0.46\\
\rule{0pt}{2ex}$^{32}$S   &0.71  &0.47    &0.44&$^{61}$Ni&1.61  &1.07    &0.63\\
\rule{0pt}{2ex}$^{33}$S   &0.62  &0.56    &0.47&$^{62}$Ni&1.57  &0.76    &0.54\\
\rule{0pt}{2ex}$^{34}$S   &0.60  &0.59    &0.45&$^{64}$Ni&1.37  &1.37    &0.91\\
\rule{0pt}{2ex}$^{36}$S   &0.75  &0.75    &0.55&$^{64}$Zn&0.50  &0.45    &0.44\\
\rule{0pt}{2ex}$^{40}$Ca&0.62  &0.44    &0.44&$^{66}$Zn&0.72  &0.64    &0.52\\
\rule{0pt}{2ex}$^{42}$Ca&0.49  &0.49    &0.46&$^{67}$Zn&0.77  &0.76    &0.56\\
\rule{0pt}{2ex}$^{43}$Ca&0.60  &0.52    &0.46&$^{68}$Zn&0.67  &0.67    &0.54\\
\rule{0pt}{2ex}$^{44}$Ca&0.59  &0.46    &0.45&$^{70}$Zn&0.57  &0.57    &0.45\\
\rule{0pt}{2ex}$^{46}$Ca&0.96  &0.96    &0.52&&&&\\
\hline
\end{tabular}}
\end{table}

\clearpage
\begin{table}
Supplementary Table~4: Core-collapse supernova yields of stable Li, Be, and B 
isotopes for the 11.8-solar-mass model with no fallback (Case 1)
\vskip 0.5cm
\centerline{\begin{tabular}{|l|c|c|c|c|}
\hline
Isotope&Yield ($M_\odot$)&Case 1&Case 2&Case 3\\
\hline 
\rule{0pt}{2ex}$^6$Li&$1.10(-11)$&$8.5(-4)$&$8.5(-4)$&$8.4(-4)$\\
\rule{0pt}{2ex}$^7$Li&$1.55(-7)$&0.84&0.83&0.82\\
\rule{0pt}{2ex}$^9$Be&$5.99(-11)$&$2.1(-2)$&$2.1(-2)$&$2.1(-2)$\\
\rule{0pt}{2ex}$^{10}$B&$1.45(-9)$&$7.6(-2)$&$7.6(-2)$&$6.8(-2)$\\
\rule{0pt}{2ex}$^{11}$B&$4.25(-7)$&5.03&5.00&4.56\\
\hline
\end{tabular}}
\end{table}

\noindent  Corresponding percentages of the solar system inventories 
are given for Case 1 and two other cases with fallback.
Note that $x(-y)$ denotes $x\times 10^{-y}$.

\clearpage
\noindent{\bf Supplementary Discussion}\\

\noindent
\underline{Core-collapse supernova (CCSN) yields of stable isotopes}.
Supplementary Table~2 gives the yields of major stable isotopes for the 
$11.8\,M_\odot$ CCSN model assuming no fallback (Case 1). As shown
in Fig.~1a of the main text, 
the yields of stable isotopes increase greatly for CCSNe of 
14--$30\,M_\odot$. For a stable isotope $^i$E of element E, the percentage 
of its solar system (SS) inventory contributed by a CCSN is
\begin{equation}
\eta(^i{\rm E})\sim\frac{fY(^i{\rm E})}{X_\odot(^i{\rm E})M_\odot}\times 100,
\end{equation}
where $Y(^i{\rm E})$ is its yield, $f$ is the fraction incorporated into each 
$M_\odot$ of the protosolar cloud, and $X_\odot(^i{\rm E})$ is its solar 
mass fraction. The CCSN contributions would introduce shifts in 
$^i$E/$^j$E, the number ratio of isotopes $^i$E and $^j$E, for SS materials.
The percentage shift for macroscopic samples can be estimated as 
$\delta(^i{\rm E}/^j{\rm E})=100[(^i{\rm E}/^j{\rm E})/(^i{\rm E}/^j{\rm E})_\odot-1]
\sim\eta(^i{\rm E})-\eta(^j{\rm E})$, where $(^i{\rm E}/^j{\rm E})_\odot$ is
the SS average.
No large shifts at the few percent level have been observed for
stable isotopes of e.g., Mg, Si, Ca, Fe, and Ni, and the observed large 
excess in $^{16}$O is most likely unrelated to its nucleosynthetic origin 
\cite{swasserburg}.

Supplementary Table~3 gives $\eta(^i{\rm E})$ 
for Case 1 of the $11.8\,M_\odot$ model without 
fallback and Cases 2 and 3 with fallback, assuming $f\sim 5\times 10^{-4}$
for all cases. It can be seen that in all cases the above CCSN contributed 
$\lesssim 1\%$ of the SS inventory for most of the major stable isotopes, 
with the highest contribution being 2.72\% for $^{18}$O. 
The shifts in $^i$E/$^j$E due to this CCSN
are entirely consistent with meteoritic constraints \cite{swasserburg}.

In Cases 2 and 3 of the $11.8\,M_\odot$ model with fallback, the percentage 
contributions to the heaviest Cr ($^{54}$Cr), Fe ($^{58}$Fe), and Ni 
($^{64}$Ni) isotopes are the highest among their respective isotopes 
(see Supplementary Table~3). This arises because fallback is 
efficient in destroying the products of explosive nucleosynthesis, but has a 
smaller effect on the neutron-rich isotopes produced by 
the slow neutron-capture ($s$) process
during pre-CCSN evolution. Macroscopic samples receiving contributions
from the $11.8\,M_\odot$ CCSN would typically show excesses of 
$^{54}$Cr/$^{52}$Cr, $^{58}$Fe/$^{56}$Fe, and $^{64}$Ni/$^{58}$Ni 
at levels of $\sim 10^{-3}$, similar to those observed in meteorites. 
As there are few satisfactory explanations of these excesses \cite{swass},
this provides additional circumstantial support for the thesis
that a low-mass CCSN with modest fallback triggered SS formation.
We note that Ref.~\cite{swass} explored a different explanation of these 
excesses by $s$-processing in asymptotic-giant-branch (AGB) stars. 
A crucial distinction between the above two explanations lie in their 
associated grains that would carry much larger anomalies. Grains from 
a low-mass CCSN would show excesses of $^{29}$Si/$^{28}$Si and 
$^{30}$Si/$^{28}$Si (see especially Case 2 in Supplementary Table~3), 
but those from AGB stars would not \cite{swass}.\\

\noindent
\underline{CCSN production of short-lived radionuclides (SLRs)}.
Most of the $^{10}$Be production occurs in the C and O shells,
where the mass fraction of $^{12}$C is high but that of $^4$He is low. A low 
$^4$He abundance is crucial in avoiding $^{10}$Be destruction via
$^{10}{\rm Be}(\alpha,n)^{13}{\rm C}$. For the same reason, neutrino-induced
production of $^{10}$Be is self-limiting because spallation of $^{12}$C and 
$^{16}$O also produces $^4$He and protons that can destroy 
$^{10}$Be via $^{10}{\rm Be}(p,\alpha)^{7}{\rm Li}$. As long as abundances 
of $^4$He and protons are low, most of the $^{10}$Be survives even when 
the production zone is heated by shock passage.
We note that Ref.~\cite{syoshida} adopted a rate for
$^{10}{\rm Be}(\alpha,n)^{13}{\rm C}$ that is orders of magnitude larger than 
currently recommended \cite{scyburt}, and therefore, greatly underestimated 
the $^{10}$Be yield. In addition, the $16.2\,M_\odot$ model used in that
work was evolved from a helium core with a fitted hydrogen envelope 
\cite{shigey} while each of our models has been evolved self-consistently
as a whole star. We find that the radii of the C/O shell in that 
$16.2\,M_\odot$ model are $\sim 2$ times larger than those in our 
$16\,M_\odot$ model. Consequently, the neutrino flux for $^{10}$Be 
production was significantly smaller in Ref.~\cite{syoshida}, which
also contributed to the much smaller yield reported there. 

The SLR $^{36}$Cl can be produced by neutrinos via 
$^{36}{\rm Ar}(\bar\nu_e,e^{+})^{36}{\rm Cl}$, where the $^{36}$Ar
is made during shock passage by explosive nucleosynthesis.
This channel accounts for $\sim 60$\%
of the $^{36}$Cl yield in our $11.8\,M_\odot$ model.
A significant fraction (up to $\sim 40$\%) of $^{26}$Al 
is made by explosive nucleosynthesis via 
$^{25}{\rm Mg}(p,\gamma)^{26}{\rm Al}$, including some enhancement
by the protons released from neutrino spallation.
$^{53}$Mn is almost entirely a product of explosive nucleosynthesis 
in all of our models.

SLRs are also produced during pre-CCSN evolution:
$^{26}$Al through hydrostatic burning at the edge of the He shell;
$^{36}$Cl, $^{41}$Ca, $^{107}$Pd, and $^{205}$Pb mainly through the 
$s$ process associated with He core burning; and 
$^{60}$Fe, $^{135}$Cs, and $^{182}$Hf mostly through the $s$ process 
associated with C burning in the O shell. The neutrons for the $s$ process 
come from $^{22}{\rm Ne}(\alpha,n)^{25}{\rm Mg}$, while the seeds come 
from the progenitor's initial composition, which is taken
as solar in all of our models.

For the $11.8\,M_\odot$ model with no fallback (Case 1), 
Supplementary Fig.~1 shows 
the fraction of the total yield of each SLR enclosed within a specific radius, 
as a function of the stellar mass enclosed within the same radius.
The left (right) vertical line indicates the boundary between
fallback and ejected material for Case 2 (3) with only 1.5\% of the 
innermost $1.02\times 10^{-2}\,M_\odot$ ($0.116\,M_\odot$) of shocked 
material ejected. As the 
enclosed fraction of the $^{10}$Be yield sharply increases near the 
boundary for Case 3, this case represents the limit of reducing 
$^{53}$Mn and $^{60}$Fe without affecting the concordance
among $^{10}$Be, $^{41}$Ca, and $^{107}$Pd shown in Fig.~2
of the main text.

Reference~\cite{stakigawa} used high-mass CCSN models with 
fallback to explain the meteoritic data on the SLRs
$^{26}$Al, $^{41}$Ca, $^{53}$Mn, and $^{60}$Fe. 
Supplementary Table~1 gives the results for our $20\,M_\odot$
and $30\,M_\odot$ models based on similar scenarios.
For the $20\,M_\odot$ ($30\,M_\odot$) model, we have used 
$f\sim 1.1\times 10^{-3}$ ($6.7\times 10^{-4}$)
and $\Delta\sim 1.2$~Myr (1.1~Myr) in equation~(1)
of the main text and assumed that a fraction $q\sim 5\times 10^{-4}$
($8\times 10^{-4}$) of the shocked material below the fallback boundary 
at the mass cut $M_{\rm cut}\sim 4.68\,M_\odot$ ($9.38\,M_\odot$)
is ejected. The corresponding parameters in Ref.~\cite{stakigawa} are
$f\sim 1.9\times 10^{-3}$ ($4.35\times 10^{-4}$), 
$\Delta\sim 1.07$~Myr (0.87~Myr), $q\sim 10^{-3}$ ($10^{-3}$),
and $M_{\rm cut}\sim 3.1\,M_\odot$ ($6.7\,M_\odot$). 
For our $20\,M_\odot$ ($30\,M_\odot$) model, contributions 
to the SS inventory of $^{16}$O, $^{17}$O, and $^{18}$O are
$\sim 3.1\%$, 1.7\%, and 21.6\% (0.3\%, 0.4\%, and 18.1\%), 
respectively. These would have caused large shifts in 
$^{18}$O/$^{16}$O and $^{17}$O/$^{18}$O that are not observed 
\cite{clayton}. Further, both our models overproduce $^{107}$Pd 
(see Supplementary Table~1 and Table~1 of the main text). 
No calculation was presented for this SLR in Ref.~\cite{stakigawa}.

With similar yields of $^{10}$Be for CCSNe of 11.8--$30\,M_\odot$,
production by these sources against decay may maintain an inventory 
of $^{10}$Be in the interstellar medium (ISM). An upper limit on the 
corresponding mass fraction can be estimated as
\begin{equation}
X_{\rm ISM}({^{10}{\rm Be}})\sim\frac{\langle Y(^{10}{\rm Be})\rangle 
R_{\rm SN}}{M_{\rm ISM}}\tau(^{10}{\rm Be})\sim 10^{-14},
\end{equation}
where $\langle Y(^{10}{\rm Be})\rangle\sim 5\times 10^{-10}\,M_\odot$
is the average yield (see Table~1 and Fig.~1b of the main text),
$\tau(^{10}{\rm Be})\sim 2$~Myr is the lifetime, and
$R_{\rm SN}\sim(10\ {\rm yr})^{-1}$, the rate of CCSNe in a total
mass $M_{\rm ISM}\sim 10^{10}\,M_\odot$ of ISM, is taken near the 
upper bound of plausible rates. The above estimate
can be combined with the mass fraction of $^9$Be in the ISM,
$X_\odot(^9{\rm Be})\sim 1.4\times 10^{-10}$, to give
($^{10}$Be/$^9$Be)$_{\rm ISM}\sim 6\times 10^{-5}$ at the time of 
SS formation. This is $\sim 10$ times less than 
the typical value in the early SS (see Table 1 of the main text). 
As we have shown, the latter can be explained
by a low-mass CCSN similar to our $11.8\,M_\odot$ model
that triggered the formation of the SS.

\noindent
\underline{CCSN remnant evolution and SLRs in the early SS}. 
The triggering of SS formation and the injection of SLRs into the early 
SS by a low-mass CCSN depend on the evolution of its remnant.
This evolution most likely occurred in a giant molecular cloud with
gas of varying density surrounding much denser clumps. The 
protosolar cloud resided in the core of one of these clumps. 
For triggering the collapse of the protosolar cloud, 
the shock velocity associated with remnant expansion should have been 
$v_\mathrm{s}\sim 20$--40~km~s$^{-1}$ \cite{sboss2010,sboss2014,sboss2015}. 
There is also a requirement on the remnant size so that a fraction 
$f\sim 5\times 10^{-4}$ of the CCSN ejecta was incorporated into each 
$M_\odot$ of the protosolar cloud to account for the SLRs 
$^{10}$Be, $^{41}$Ca, and $^{107}$Pd simultaneously. 
For a remnant with a shock radius $R_\mathrm{s}$ colliding
with a cloud of radius $r_\mathrm{c}$, only a fraction $\sim r_\mathrm{c}^2/(4R_\mathrm{s}^2)$ of 
the remnant material would be available for injection. So $f$ can be 
estimated as
\begin{equation}
f\sim\epsilon_{\rm in}\left(\frac{r_\mathrm{c}^2}{4R_\mathrm{s}^2}\right)
\sim 2.5\times 10^{-4}\left(\frac{\epsilon_{\rm in}}{0.1}\right)
\left(\frac{r_\mathrm{c}}{0.1\ {\rm pc}}\right)^2\left(\frac{\rm pc}{R_\mathrm{s}}\right)^2,
\end{equation}
where $\epsilon_{\rm in}$ is the efficiency of injecting the relevant part of 
the remnant material into each $M_\odot$ of the protosolar cloud. 
Guided by simulations \cite{sboss2010,sboss2014,sboss2015}, we take
$\epsilon_{\rm in}\sim 0.1$ for a typical clump core of $\sim 1\,M_\odot$ 
with $r_\mathrm{c}\sim 0.1$~pc \cite{cloud}. Based on the above discussion, 
our proposed trigger scenario requires that the low-mass CCSN remnant 
must have had $v_\mathrm{s}\sim 20$--40~km~s$^{-1}$ and $R_\mathrm{s}\sim 1$~pc when colliding 
with the protosolar cloud. In addition, these conditions must have been reached 
within $\Delta\sim 1$~Myr to account for the pertinent SLRs, especially 
$^{41}$Ca with a very short lifetime of $\sim 0.15$~Myr.

The shock velocities 
of concern typically occur when the remnant is in the pressure-driven 
snowplow (PDS) phase. For reference, we consider the simple case
of a CCSN remnant expanding in a uniform ISM. At the onset of the PDS 
phase, the shock radius $R_{\rm PDS}$ and velocity $v_{\rm PDS}$ 
\cite{cioffi} are approximately given by
\begin{eqnarray}
R_{\rm PDS}&\sim&1.01\left(\frac{E}{10^{50}\ {\rm erg}}\right)^{2/7}
\left(\frac{100\ {\rm cm}^{-3}}{n_0}\right)^{3/7}\,{\rm pc},\\
v_{\rm PDS}&\sim&676\left(\frac{E}{10^{50}\ {\rm erg}}\right)^{1/14}
\left(\frac{n_0}{100\ {\rm cm}^{-3}}\right)^{1/7}\,{\rm km~s}^{-1},
\end{eqnarray}
where $E$ is the explosion energy of the CCSN, and $n_0$ is
the number density of hydrogen atoms in the ISM.
The remnant evolution during the PDS phase \cite{cioffi} is 
approximately described by
\begin{eqnarray}
R_\mathrm{s}&\sim&R_{\rm PDS}\left(\frac{4}{3}t_*-\frac{1}{3}\right)^{3/10},\\
v_\mathrm{s}&\sim&v_{\rm PDS}\left(\frac{4}{3}t_*-\frac{1}{3}\right)^{-7/10},
\end{eqnarray}
where $t_*\equiv t/t_{\rm PDS}$ is the time $t$ since the explosion
in units of
\begin{equation}
t_{\rm PDS}\sim 584\left(\frac{E}{10^{50}\ {\rm erg}}\right)^{3/14}
\left(\frac{100\ {\rm cm}^{-3}}{n_0}\right)^{4/7}\,{\rm yr}.
\end{equation}

Using $E\sim 10^{50}$~erg for the low-mass CCSN and 
$n_0\sim 100$~cm$^{-3}$ for a typical giant molecular cloud 
\cite{cloud}, we find that for the above simple case, the remnant 
reaches $v_\mathrm{s}\sim 40$~km~s$^{-1}$ at $t\sim 2.5\times 10^4$~yr. 
The corresponding $R_\mathrm{s}\sim 3.4$~pc gives $f\sim 2\times 10^{-5}$.
This should be regarded as a lower limit on $f$ because the shock 
wave can be slowed down more efficiently in a giant molecular cloud 
with dense clumps . For example, if at the onset of the PDS phase
the remnant in the simple case encounters a clump with a hydrogen 
density of $n_0'\sim 2\times10^3$~cm$^{-3}$ and a radius of 
$\sim 1$~pc \cite{cloud}, then by momentum conservation relevant 
for the PDS phase, the shock wave approaches the core of the clump 
with $v_\mathrm{s}\sim (n_0/n_0')v_{\rm PDS}\sim 34$~km~s$^{-1}$ but its effective
radius remains close to $R_\mathrm{s}\sim 1$~pc. In this example, the
conditions for triggering the collapse of the core and injecting
SLRs into it would be satisfied. Based on the above discussion, 
we consider our trigger scenario reasonable and urge that
simulations of remnant evolution in a giant 
molecular cloud be carried out to provide more rigorous results.
We note that the time of remnant expansion must have been 
far shorter than $\Delta\sim 1$~Myr. Therefore, this interval must 
reflect the timescales associated with the collapse of the protosolar
cloud and the formation of the first solids in the early SS.

Evolution of a CCSN remnant prior to the PDS phase is associated 
with acceleration of cosmic rays (CRs), which can produce $^{10}$Be 
\cite{sepi}. We consider the amount of $^{10}$Be produced by CRs 
inside the remnant up to the onset of the PDS phase and compare
this to the low-mass CCSN yield. 
Using $n_0\sim 100$~cm$^{-3}$ but an explosion energy 10 
times too high for the low-mass CCSN, Ref.~\cite{sepi} found
that CRs can produce $^{10}$Be/$^9$Be~$\sim 2.5\times 10^{-3}$ at 
the maximum. Adopting this upper limit and a total mass of swept-up 
ISM
\begin{equation}
M_{\rm PDS}\sim 10\,M_\odot\left(\frac{n_0}{100\ {\rm cm}^{-3}}\right)
\left(\frac{R_{\rm PDS}}{\rm pc}\right)^3,
\end{equation}
we estimate that CRs can
produce at most $\sim 4\times 10^{-12}\,M_\odot$ of $^{10}$Be, which 
is far below the low-mass CCSN yield of 
$\sim 3.26\times 10^{-10}\,M_\odot$ (see Table~1 of the main text). 
Therefore, even allowing for complications of remnant evolution, 
CR production inside the remnant would have been a subdominant 
contribution to the $^{10}$Be in the early SS. 
 
Reference~\cite{sepi} considered a remnant interacting with the protosolar 
cloud and suggested that CR production of $^{10}$Be inside the cloud 
might have provided this SLR to the calcium-aluminum-rich inclusions 
with Fractionation and Unidentified Nuclear isotope effects (FUN-CAIs).
However, it is not clear how this production actually took place when
the very small size of the cloud relative to the remnant is taken into
account. It is highly desirable to extend the study in Ref.~\cite{sepi} to
our proposed low-mass CCSN trigger scenario.\\

\noindent
\underline{Potential tests for a low-mass CCSN trigger: Li, Be, B}. 
In our proposed scenario, a low-mass CCSN trigger provided the bulk of 
the $^{10}$Be inventory in the early SS as indicated by canonical CAIs. 
CR production associated with the CCSN remnant might have provided 
$^{10}$Be to FUN-CAIs \cite{sepi}. Any $^{10}$Be production by CRs and
solar energetic particles (SEPs) \cite{sgou1,sgou2} would be in addition to 
the injection from the CCSN but generally at subdominant levels 
consistent with the observed variations of $^{10}$Be/$^9$Be in 
canonical CAIs \cite{smckeegan,smacpherson,swielandt,ssrin}.

We propose a potential test of the above scenario based on the distinct
yield pattern of Li, Be, and B isotopes for the CCSN. Supplementary
Table~4 gives the yields of $^6$Li, $^7$Li, $^9$Be, $^{10}$B, and
$^{11}$B for the $11.8\,M_\odot$ model with no fallback (Case 1). 
The fallback in Cases 2 and 3 causes little change in these results.
It can be seen that the CCSN predominantly produces $^7$Li and
$^{11}$B, which is a feature of neutrino-induced nucleosynthesis
\cite{nuprocess}. This is in sharp contrast to the production by
CRs or SEPs with much higher energy than CCSN neutrinos.
For example, Ref.~\cite{sepi} gave relative number yields of
$^6{\rm Li}:{^7{\rm Li}}:{^9{\rm Be}}:{^{10}{\rm B}}:{^{11}{\rm B}}
\sim 1:1.8:0.11:0.43:1.1$.

The presence of $^{10}$Be in the early SS is established 
by the correlation between $^{10}$B/$^{11}$B and $^9$Be/$^{11}$B,
from which the initial values ($^{10}$Be/$^9$Be)$_0$ and
$(^{10}$B/$^{11}$B)$_0$ at the time of $^{10}$Be incorporation 
are obtained. In our scenario, the low-mass CCSN trigger provided 
the bulk of the $^{10}$Be and $\sim 5\%$ of the $^{11}$B in the SS 
(see Supplementary Table~4). Consequently, we expect that samples 
with higher ($^{10}$Be/$^9$Be)$_0$ would have lower 
($^{10}$B/$^{11}$B)$_0$ due to the excess of $^{11}$B over $^{10}$B
that accompanied the $^{10}$Be from the CCSN. The variations 
in ($^{10}$B/$^{11}$B)$_0$ should be at the level of $\sim 5\%$. 
Such variations are consistent with the data reported in
Refs.~\cite{swielandt,ssrin}. It remains to be seen if future meteoritic 
studies with more samples and better precision can establish the above
relationship rigorously, thereby providing a test for the low-mass
CCSN trigger.

A correlation between $^7$Li/$^6$Li and $^9$Be/$^6$Li was
found in a sample with 
($^{10}$Be/$^9$Be)$_0=(8.8\pm0.6)\times 10^{-4}$ and
interpreted as indicating the presence of the SLR $^7$Be 
in the early SS \cite{li1}. If true, the extremely short lifetime of 
77~days for $^7$Be would almost certainly require irradiation
by SEPs for its production, and by association, the same 
mechanism may also have produced the $^{10}$Be in the sample. 
However, the above result was disputed and the controversy 
remains unresolved \cite{li2,li3}. Here we propose an alternative
explanation for the tantalizing correlation between $^7$Li/$^6$Li 
and $^9$Be/$^6$Li. We note that the low-mass CCSN trigger
also provided $\sim 0.8\%$ of the $^7$Li in the SS but
a negligible amount of $^6$Li. We expect that portions of
a sample that received higher amounts of $^{10}$Be from the
CCSN would also have higher amounts of $^7$Li. For
a uniform ($^{10}$Be/$^9$Be)$_0$ across the sample, the
above relationship would translate into an apparent correlation 
between $^7$Li/$^6$Li and $^9$Be/$^6$Li, which nevertheless
has nothing to do with the presence of $^7$Be in the early SS.
This explanation can be tested more directly by checking
the relationship between $^7$Li/$^6$Li and 
($^{10}$Be/$^9$Be)$_0$ for a wide range of samples.
In our scenario, $^7$Li/$^6$Li should increase with 
($^{10}$Be/$^9$Be)$_0$. As only a relatively small amount
of $^7$Li was added by the low-mass CCSN, high-precision
measurements are required to check this relationship.
Such measurements would provide an additional test for the 
low-mass CCSN trigger and also help resolve the controversy 
over the $^7$Be result.

\clearpage
\noindent{\bf Supplementary References}


\end{document}